\documentclass[aps,prc,twocolumn,groupedaddress,showpacs,floatfix]{revtex4}
\usepackage{graphicx}
\usepackage{dcolumn}
\usepackage{bm}

\newcommand{\raa}{($\alpha$,$\alpha$)}

\newcommand{\rag}{($\alpha$,$\gamma$)}

\newcommand{\ran}{($\alpha$,n)}
\newcommand{\rlid}{($^6$Li,d)}

\newcommand{\al}{$\alpha$}

\begin{document}

{
\title{
$\mathbf\alpha$-cluster states in intermediate mass nuclei
}

\author{Peter Mohr\footnote{Electronic address: WidmaierMohr@compuserve.de}}
\affiliation{
Diakonie-Klinikum Schw\"abisch Hall, D-74523 Schw\"abisch Hall,
Germany
}

\date{\today}

\begin{abstract}
Properties of intermediate mass nuclei have been investigated within
the framework of the $\alpha$-cluster model in combination with
systematic double-folding potentials. Previously, this
$\alpha$-cluster model has been widely applied to light nuclei, in
particular to $^8$Be = $\alpha$ $\otimes$ $\alpha$, $^{20}$Ne =
$^{16}$O $\otimes$ $\alpha$, and $^{44}$Ti = $^{40}$Ca $\otimes$
$\alpha$, and to heavy nuclei, in particular to $^{212}$Po =
$^{208}$Pb $\otimes$ $\alpha$.  In the present work a wide range of
nuclei is investigated with the magic neutron number $N = 50$ in the
mass range around $A \approx 80 - 100$: ($A$+4,$N$=52) = ($A$,$N$=50)
$\otimes$ $\alpha$. It is found that excitation energies, decay
properties, and transition strengths can be described successfully
within this model. The smooth and small variation of the underlying
parameters of the $\alpha$-nucleus potential may be used for
extrapolations to predict experimentally unknown properties in the
nuclei under study.
\end{abstract}

\pacs{21.60.Gx,27.50.+e,27.60.+j}

\maketitle
}

\section{\label{sec:intro}Introduction}
Atomic nuclei are complex many-body systems. Their properties are
defined by the short-range nuclear interaction and the long-range
Coulomb interaction which have to be included in the
quantum-mechanical many-body Schroedinger equation. The many-body
problem may be dramatically simplified in some cases where the nucleus
can be considered to be composed of two inert clusters like an
$\alpha$-particle and a closed-shell nucleus. In these cases a simple
two-body model is able to reproduce many properties like e.g.\
excitation energies and decay properties -- provided that the
interaction potential can be well described. This $\alpha$-cluster
model has been widely applied to light nuclei, in particular to $^8$Be
= $\alpha$ $\otimes$ $\alpha$, $^{20}$Ne = $^{16}$O $\otimes$
$\alpha$, and $^{44}$Ti = $^{40}$Ca $\otimes$ $\alpha$, and to heavy
nuclei, in particular to $^{212}$Po = $^{208}$Pb $\otimes$
$\alpha$. Here I apply this $\alpha$-cluster model in combination with
systematic $\alpha$-nucleus double-folding potentials to the analysis
of a wide range of nuclei with the magic neutron number $N = 50$ in
the mass range around $A \approx 80 - 100$: ($A$+4,$N$=52) =
($A$,$N$=50) $\otimes$ $\alpha$. This study extends earlier work which
has focused on $^{94}$Mo = $^{90}$Zr $\otimes$ $\alpha$.

The \al -particle is the lightest doubly-magic nucleus. Thus it is
strongly bound, and its first excited state is located at the high
excitation energy $E_x = 20.2$\,MeV \cite{ENSDF,Til92}. Although
composed of four nucleons, the \al -particle may be considered as
inert in many investigations, i.e.\ the internal structure of the \al
-particle may be neglected. It has been stated that this ``concept of
\al -clustering is essential for understanding the structure of light
nuclei'' \cite{Ohk98}, and this concept has been extended to heavy nuclei
(e.g.\ \cite{Mic98,Yam98,Sak98,Ueg98,Has98,Koh98,Toh98}).

As a consequence of the strong binding of \al -particles many
observables in the interaction of two \al -particles can be described
successfully using a simple two-body model. Typical observables are
scattering cross sections of the $^4$He\raa $^4$He reaction and
excitation energies, transition probabilities, and decay properties of
the $^8$Be = \al\ $\otimes$ \al\ nucleus. A prerequisite for the
successful application of the simple two-body model is an effective
potential between the two \al -particles. It has been shown that the
double-folding potential and the widely used DDM3Y interaction provide
an excellent description of the above observables \cite{Mohr94}.

Similar arguments hold for the description of nuclei which are
composed of another doubly-magic nucleus and an \al -particle, and
numerous studies have been devoted to $^{20}$Ne = $^{16}$O $\otimes$
\al , $^{44}$Ti = $^{40}$Ca $\otimes$ \al , and $^{212}$Po =
$^{208}$Pb $\otimes$ \al\ (see e.g.\ the review papers of the
dedicated special issue of Prog.\ Theor.\ Phys.\ Suppl.\ {\bf 132},
\cite{Ohk98,Mic98,Yam98,Sak98,Ueg98,Has98,Koh98,Toh98}). However,
there is no stable doubly-magic nucleus with the magic neutron number
$N = 50$. In the present work I analyze nuclei with ($A$+4,$N$=52) =
($A$,$N$=50) $\otimes$ \al\ between the two instable $N = 50$
doubly-magic nuclei $^{78}$Ni ($Z = 28$) and $^{100}$Sn ($Z =
50$). Previous studies (e.g.\ \cite{Mic98,Ohk95,Buck95}) have been
restricted to $^{94}$Mo = $^{90}$Zr $\otimes$ \al\ because of the
subshell closure at $Z = 40$ between $f$ and $p$ subshells and the
proton $g_{9/2}$ subshell. It will be shown that there is no need for
this restriction. The proton subshell closure at $Z = 40$ turns out to
be not important for the properties of \al -cluster states. Instead,
the \al -cluster model can be applied successfully in the same way to
all nuclei under study except the semi-magic $Z = 50$ nucleus
$^{102}$Sn = $^{98}$Cd $\otimes$ \al . The present study is restricted
to even-even nuclei. An extension of the present study to even-odd
nuclei is possible e.g.\ for $^{93}$Nb = $^{89}$Y $\otimes$ \al\ which
will be published together with the analysis of the $^{89}$Y\raa
$^{89}$Y scattering cross section \cite{Kiss08}.

The paper is organized as follows. In Sect.~\ref{sec:model} the
ingredients of the simple two-body model are briefly summarized. In
particular, the underlying double-folding potential and the formalism
for the calculation of bound state properties and transition strengths
is presented. In Sect.~\ref{sec:results} the results of the
calculations, i.e.\ the systematic behavior of the potential
parameters for the description of the bound state energies, are
shown. Sect.\ \ref{sec:disc} gives a discussion of the results, and
finally conclusions are drawn in Sect.~\ref{sec:conc}. In the
following discussion proton (neutron, mass) numbers $N_C$ ($Z_C$,
$A_C$) of the compound nucleus ($A_C$=$A$+4,$N_C$=$N$+2) = ($A$,$N$)
$\otimes$ \al\ are indexed by the subscript $C$ whereas $Z$ ($N$, $A$)
without index refers to the core nucleus; e.g., properties of the $N_C
= 52$, $Z_C = 42$, $A_C = 94$ nucleus $^{94}$Mo are calculated from
the potential between the $N = 50$, $Z = 40$, $A = 90$ core $^{90}$Zr
$\otimes$ \al .

\section{\label{sec:model}Ingredients of the model}

\subsection{\label{sec:fold}Folding potentials}
The basic ingredient of the present study is the \al -nucleus
potential. Various parametrizations have been used in
literature. However, recent systematic studies have concentrated on
folding potentials and Woods-Saxon potentials. It has been shown that
the parameters of folding potentials, in particular the volume
integral $J_R$, show a very systematic behavior for intermediate and
heavy mass nuclei \cite{Atz96}. This study \cite{Atz96} has been
extended to the analysis of elastic scattering data at astrophysically
relevant energies around the Coulomb barrier
\cite{Mohr97,Ful01,Gal05,Kiss06,Kiss07}.
Experimental data for \rag\ capture reactions and other
\al -induced reactions like \ran\ have been studied using folding
potentials and other potential parametrizations, see e.g.\ in
\cite{Dem02,Gyu06,Som98,Spy07}. Further information on the systematics
of folding potentials has been obtained from the analysis of \al
-decay data for superheavy nuclei (e.g.\
\cite{Mohr06,Zha07,Sam07,Xu07,Xu05,Cho05,Den05}), for neutron-deficient
$p$-nuclei \cite{Mohr00,Fuj02,Xu05b}, and for nuclei slightly above the
doubly-magic $N = Z = 50$ nucleus $^{100}$Sn \cite{Xu06,Mohr07}. The
present work extends the study of \cite{Mohr07} to lighter $N =
50$ nuclei because the double-folding potentials have proven to be
reliable in such a broad range of masses.

Global and local Woods-Saxon potentials have also been determined
succssfully for the mass range under study in
\cite{Avr06,Avr06b,Avr03}. The description of scattering data is
similar to folding potentials \cite{Kiss07}. However, as will be shown
in Sect.~\ref{sec:bound}, it is not possible to obtain a reasonable
description of the excitation energies of rotational bands in
($A$+4,$N$=52) = ($A$,$N$=50) $\otimes$ \al\ nuclei.

Other parametrizations of the \al -nucleus potential like e.g.\
modified Woods-Saxon potentials \cite{Buck95} or the so-called
{\it{cosh}} potential \cite{Buck93} are not analyzed in this
work. Although the {\it{cosh}} potential was able to describe \al
-decay properties, the {\it{cosh}} potential or other potentials have
not been used for the simultaneous description of scattering and
reaction cross sections and bound state and decay properties.

The double-folding potential $V_F(r)$ is calculated from the densities
of the interacting nuclei and an effective nucleon-nucleon
interaction. In the present work the nuclear densities were derived
from electron scattering data which are compiled in \cite{Vri87}. Many
nuclei under study are unstable, and electron scattering data are not
available in \cite{Vri87}. Therefore, the densities for all $N = 50$
nuclei in the present study were derived in the same way as in
\cite{Mohr06,Mohr07}; there it was shown that an averaged
two-parameter Fermi distribution provides reasonable densities over a
broad mass range by a simple scaling of the radius parameter $r \sim
A^{1/3}$. As will be discussed later, the resuls do not show a strong
dependence on the radius parameter. The widely used DDM3Y interaction
is also applied in this work. Details on the calculation of the
double-folding potential and the effective interaction can be found in
\cite{Sat79,Kob84,Abe93,Atz96}.

The total potential $V(r)$ is given by the sum of the nuclear
potential $V_N(r)$ and the Coulomb potential $V_C(r)$:
\begin{equation}
V(r) = V_N(r) + V_C(r) = \lambda \, V_F(r) + V_C(r)
\label{eq:vtot}
\end{equation}
The Coulomb potential is taken in the usual form of a homogeneously
charged sphere where the Coulomb radius $R_C$ has been chosen
identically with the root-mean-square radius $r_{\rm{rms}}$ of the
folding potential $V_F$. The folding potential $V_F$ is scaled by
a strength parameter $\lambda$ which is of the order of $1.0 -
1.3$. This leads to volume integrals of about $J_R \approx
300$\,MeV\,fm$^3$ for all nuclei under study and is in agreement with
systematic \al -nucleus potentials derived from elastic scattering
\cite{Atz96,Dem02,Avr03}. (Note that as usual the negative sign of
$J_R$ is omitted in this work.)

For comparison, calculations have also been performed with nuclear
potentials $V_N$ of Woods-Saxon shape: $V_N(r) = V_0 \times [1 +
\exp{(r-R)/a}]^{-1}$ with the potential depth $V_0$, radius parameter $R =
r_0 \times A^{1/3}$ and diffuseness $a$.

It has been suggested that a temperature dependence of the optical
potential may improve the simultaneous description of scattering,
reaction, and decay data \cite{Avr06b}. The analysis of scattering and
reaction data requires a complex optical potential where the imaginary
part describes the absorption into other channels. The present study
focuses on bound state properties which can be calcuated from the real
part of the potential, see Eq.~(\ref{eq:vtot}). Because of the energy
and density dependence of the interaction and because of the above
mentioned temperature dependence it is clear that the real potentials
from this study will require minor modification so that they can be
used as real part of a complex potential in the analysis of
scattering and reaction data.

\subsection{\label{sec:bound}Bound and quasi-bound states}
From a given nuclear potential it is a straightforward task to
calculate the eigenstates of the Hamilton operator, i.e.\ the energies
$E$ and wave functions $u(r)$. The Pauli principle is taken into
account by the so-called Wildermuth condition which relates the
quantum numbers $Q, N, L$ of the \al -particle to the quantum numbers
$q_i, n_i, l_i$ of the four constituent nucleons:
\begin{equation}
Q = 2N + L = \sum_{i=1}^4 (2n_i + l_i) = \sum_{i=1}^4 q_i
\label{eq:wild}
\end{equation}
where $Q$ is the number of oscillator quanta, $N$ is the number of
nodes and $L$ the relative angular momentum of the \al -core wave
function, and $q_i = 2n_i + l_i$ are the corresponding quantum numbers
of the four nucleons forming the \al\ cluster. I have taken $q = 4$
for the two neutrons above the neutron number $N=50$ and $q = 4$ ($q =
3$) for protons above (below) the proton number $Z = 40$. This leads
to $Q = 16$ for nuclei above $^{94}$Mo = $^{90}$Zr $\otimes$ \al\ and
$Q = 14$ for nuclei below $^{92}$Zr = $^{88}$Sr $\otimes$ \al .

\begin{figure}[htb]
\includegraphics[ bb = 50 60 480 345, width = 80 mm, clip]{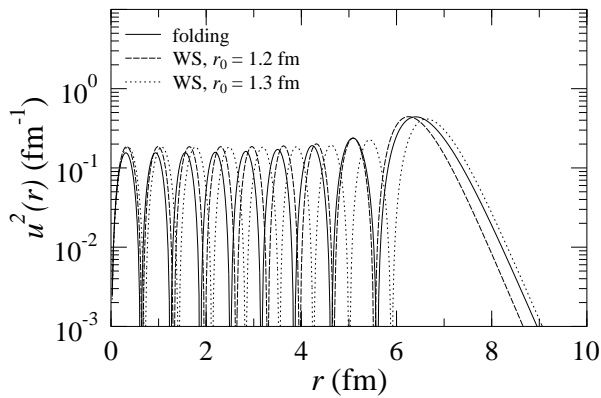}
\caption{
  \label{fig:wave} 
  Square of the wave function $u(r)$ of the $0^+$ ground state of $^{96}$Ru
  = $^{92}$Mo $\otimes$ \al\ at $E = -1692$\,keV using the folding
  potential with $\lambda = 1.1965$ and the Woods-Saxon potentials
  with two different radius parameters $r_0 = 1.2$\,fm and
  1.3\,fm. Corresponding to $Q = 16$, the number of nodes is eight.  
}
\end{figure}

In a first calculation the strength of the nuclear potential is
adjusted to reproduce the binding energy $E_B$ of the \al -particle to
the $N=50$ core. E.g., for the $0^+$ ground state wave function of the
nucleus $^{96}$Ru = $^{92}$Mo $\otimes$ \al\ one finds $E_B =
-1692$\,keV \cite{ENSDF} ($E_B < 0$ for bound states). For the folding
potential a strength parameter of $\lambda = 1.1965$ is required to
reproduce this energy with $Q = 16$, i.e., a wave function with
angular momentum $L = 0$ and $N = 8$ nodes. For the Woods-Saxon
potential one finds $V_0 = 142.97$\,MeV (162.38\,MeV) using geometry
parameters of $r_0 = 1.3$\,fm (1.2\,fm) and $a = 0.7$\,fm. These
geometry parameters of the Woods-Saxon potential are close to the
Woods-Saxon potentials derived from scattering data
\cite{Avr06,Avr06b,Avr03}. The square of the wave functions $u(r)$ in
the different potentials is shown in Fig.~\ref{fig:wave}.

\begin{figure}[htb]
\includegraphics[ bb = 125 40 525 830, width = 70 mm, clip]{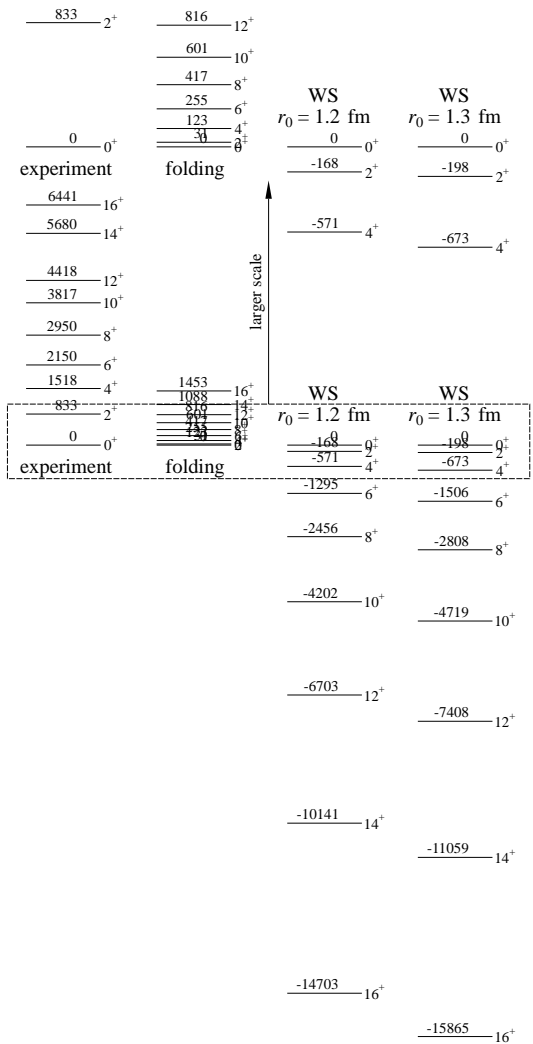}
\caption{
  \label{fig:exc} 
  Excitation energies for the nucleus $^{96}$Ru = $^{92}$Mo $\otimes$
  \al . Experimentally one finds a rotational band $0^+$, $2^+$,
  $4^+$, $\ldots$ $16^+$ (left). Although compressed in energy, the
  folding potential reproduces such a rotational band (with the
  energies $E_x = 31$, 123, 255, 417, 601, 816, 1088, and 1453\,keV
  for the $2^+$, $4^+$, $\ldots$ $16^+$ states). In contrast, the Woods-Saxon
  potentials show an inversion of the levels (right). All energies are given
  in keV. The low-enery region (as indicated by the dashed box) is
  scaled up in the upper part of the diagram for better readability.
}
\end{figure}

In the next step the energies of all excited states with $Q = 16$ are
calculated using exactly the same potential as for the ground
state. The excitation energies $E_x$ are defined by $E = E_B +
E_x$. The result is shown in Fig.~\ref{fig:exc} for the chosen example
of $^{96}$Ru = $^{92}$Mo $\otimes$ \al . Experimentally states with
quantum numbers from $0^+$ to $16^+$ are known which form a rotational
band although the energies do not follow exactly the rigid rotator
rule $\sim L(L+1)$. A similar rotational behavior is found for the
folding potential; however, the excitation energies are much lower
than the experimental values. In contrast, the Woods-Saxon potentials
show an inversion of the levels; i.e., the $16^+$ state is strongest
bound in the Woods-Saxon potential. This finding is independent of
details of the chosen geometry parameters. In Fig.~\ref{fig:exc}
results for two radius parameters $r_0 = 1.2$\,fm and 1.3\,fm are
shown. Whereas in the case of the folding potential a minor
readjustment of the potential strength of less than 5\,\% is
sufficient to reproduce the excitation energies (see
Sect.~\ref{sec:results}), the Woods-Saxon potential requires strong
modification of more than 20\,\% to reproduce the excitation energy
spectrum.

It is interesting to note that the inversion of the excitation
energies in the Woods-Saxon potential is not directly related to the
width of the potential. As can be seen from Fig.~\ref{fig:wave}, the
wave function of the Woods-Saxon potential with $r_0 = 1.2$\,fm is
concentrated at smaller radii than the folding wave function whereas
the wave function in the Woods-Saxon potential with $r_0 = 1.3$\,fm is
concentrated at larger radii. Nevertheless, both Woods-Saxon
potentials show the inversion of excitation energies, whereas the
folding potential reproduces a rotational band (see also
Fig.~\ref{fig:exc}).

\subsection{\label{sec:trans}Transition strengths}
Reduced transition strengths $B(E{\cal{L}})$ for electromagnetic
transitions $L_i \rightarrow L_f$ can be calculated from the bound
state wave functions $u_{L_i}(r)$ and $u_{L_f}(r)$. The transition
strengths scale with the square of the overlap integral
\begin{equation}
B(E{\cal{L}}) \sim \left| \int u_{L_f}(r) \, r^{\cal{L}} \, u_{L_i}(r)
\, dr \right|^2
\label{eq:BE}
\end{equation}
The full formalism for the calculation of $B(E{\cal{L}})$ values can
be found e.g.\ in \cite{Hoy94,Buck94,Buck77}. The present study will
be restricted to quadrupole transitions with ${\cal{L}} = 2$. One
expects enhanced transition strengths of the order of several
Weisskopf units (W.u.) for the intraband transitions within a
rotational band. The present study extends earlier work which has
focused on transitions in $^{94}$Mo = $^{90}$Zr $\otimes$ \al\
\cite{Mic98,Ohk95,Buck95}.

\begin{figure}[htb]
\includegraphics[ bb = 40 65 480 465, width = 80 mm, clip]{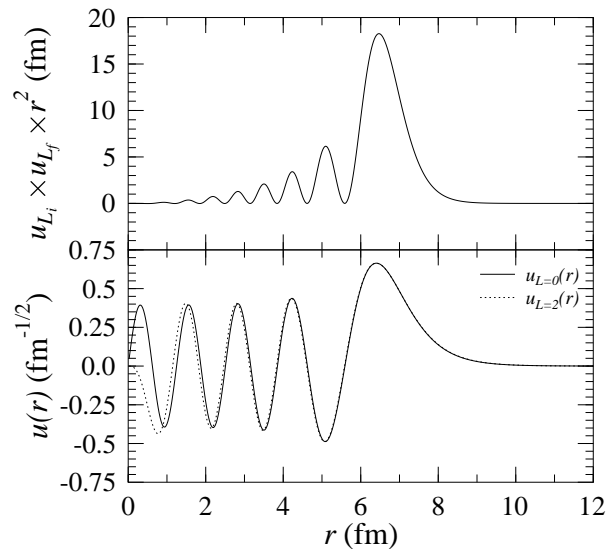}
\caption{
  \label{fig:BE} 
  Wave functions $u_{L=0}(r)$ and $u_{L=2}(r)$ (lower part), and the
  integrand of the overlap integral in Eq.~(\ref{eq:BE}) (upper part)
  for the transition from the $2^+$ state at $E_x = 833$\,keV to the
  $0^+$ ground state in $^{96}$Ru.  
}
\end{figure}

As an example, Fig.~\ref{fig:BE} shows the wave functions
$u_{L=0}(r)$, $u_{L=2}(r)$, and the integrand of the overlap integral
in Eq.~(\ref{eq:BE}) for the transition from the $2^+$ state at $E_x =
833$\,keV to the $0^+$ ground state in $^{96}$Ru. The dominating
contribution of the integral in Eq.~(\ref{eq:BE}) is located at the
nuclear surface around $6 - 8$\,fm. 

It is interesting to note that the calculated transition strengths
$B(E{\cal{L}})$ show only a weak dependence on the excitation energy
$E_x$ wich is discussed now for the shown example of the $2^+
\rightarrow 0^+$ transition in $^{96}$Ru. In a first calculation
$B(E{\cal{L}})$ is calculated using $\lambda = 1.1965$ (adjusted to
the binding energy of the ground state). This leads to $E_x(2^+) =
31$\,keV (as shown in Fig.~\ref{fig:exc}) instead of the experimental
$E_x = 833$\,keV. In a second calculation the wave function of the
$2^+$ state is calculated using a slightly changed potential strength
$\lambda = 1.1852$ (adjusted to fit the excitation energy of the $2^+$
state). The resulting $B(E{\cal{L}})$ value changes only by about
2\,\% between these two calculations which can be explained by
Figs.~\ref{fig:wave} and \ref{fig:BE}. The dominating contribution of
the integral in Eq.~(\ref{eq:BE}) comes from the nuclear interior and
surface. However, a change in the excitation energy mainly leads to a
change in the asymptotic behavior of the wave function, i.e.\ a
different slope of the wave function in the exterior region. Although
the dependence of the $B(E{\cal{L}})$ values on the excitation energy
is relatively small, all $B(E{\cal{L}})$ values are calculated from
wave functions with correct asymptotic bahavior, i.e.\ the potential
strength has been readjusted to fit the respective excitation energy
$E_x$, see Sect.\ \ref{sec:results} and Fig.~\ref{fig:laml}.

\section{\label{sec:results}Results}
It is one aim of the present investigation to study the behavior of
the potential strength parameter $\lambda$ of the folding potential
for a broad range of $N = 50$ nuclei. As will be shown, the parameter
$\lambda$ shows a very systematic and regular behavior. This can be
used to predict up-to-now unknown properties like e.g. \al -decay
energies or excitation energies \cite{Mohr07}. Alternatively,
clear deviations from the systematic behavior of the potential strength
parameter $\lambda$ may be interpreted as indications for shell
closures \cite{Mohr06}.

\begin{figure}[htb]
\includegraphics[ bb = 50 65 470 410, width = 80 mm, clip]{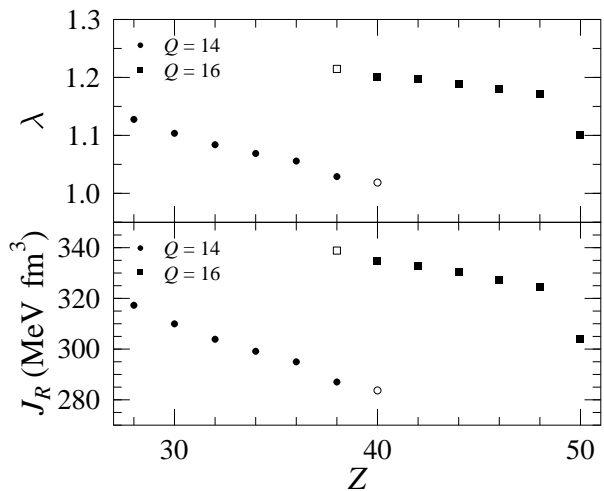}
\caption{
  \label{fig:lamz} 
  Potential strength parameter $\lambda$ (upper part) and volume
  integrals $J_R$ (lower part) for nuclei ($A$+4,$N$=52) =
  ($A$,$N$=50) $\otimes$ \al\ in dependence of the proton number
  $Z$. Data for $Z = 50$ are taken from the extrapolation in
  \cite{Mohr07}. Further discussion see text.  
}
\end{figure}

The variation of the potential strength parameter $\lambda$ and the
resulting volume integral $J_R$ are shown in Fig.~\ref{fig:lamz} for
$N=50$ nuclei from $^{78}$Ni up to $^{100}$Sn. The strength parameter
$\lambda$ has been adjusted to reproduce the binding energies of the
nuclei from $^{82}$Zn = $^{78}$Ni $\otimes$ \al\ to $^{104}$Te =
$^{100}$Sn $\otimes$ \al\ which are taken from \cite{ENSDF,Wap03}.
It is obvious from Fig.~\ref{fig:lamz}
that both diagrams look very similar. Thus, in the following
presentation of the results only the systematics of the strength
parameter $\lambda$ is discussed. A similar systematics is obtained
for the volume integrals $J_R$. Note that minor differences for the
potential strength parameter $\lambda$ for $^{94}$Mo = $^{90}$Zr
$\otimes$ \al\ between this study and earlier work in
\cite{Mic98,Ohk95} are the consequence of the global parametrization
for the nuclear density used in this study.

The variation of $\lambda$ in dependence of the proton number $Z$ in
Fig.~\ref{fig:lamz} is very smooth except around $Z=40$ and
$Z=50$. Both discontinuities are related to shell closures. Changes of
the volume integral $J_R$ by more than about 10\,MeV\,fm$^3$
(corresponding to changes in $\lambda$ by more than about 0.035) are
typical for the crossing of a shell closure \cite{Mohr06}. Usually,
the crossing of a shell closure should be combined with a change of
the oscillator quanta $Q$. However, the situation around the magic
number $Z = N = 50$ is complicated by the fact that the $g_{9/2}$
subshell ($q = 4$) is located at relatively low energies close to the
$p$ and $f$ subshells with $q = 3$.

From $Z = 48$ to $Z=50$ the oscillator quanta $Q$ in the model do not
change. However, the strength parameter $\lambda$ changes by about
0.07 which reflects the strong $Z = 50$ shell closure at
$^{100}$Sn. (The potential parameters for $^{100}$Sn have been derived
from the systematics of various tin isotopes \cite{Mohr07}.) Such
changes in the potential strength parameter $\lambda$ may be used to
assign unknown magic numbers, e.g.\ for superheavy nuclei \cite{Mohr06}.

The discontinuity at the subshell closure at $Z = 40$ between the $f$
and $p$ subshells and the proton $g_{9/2}$ subshell turns out to be an
artefact. Below $Z=40$, the wave functions have been calculated with
$Q = 14$, i.e.\ seven nodes for the $L=0$ ground state wave
function. Above $Z = 40$, the node number has been increased by one to
eight nodes ($Q = 16$). Obviously, the potential strength has to be
increased significantly to obtain a wave function with an additional
node at similar energies. As will be shown in the next paragraph, the
change in the potential strength parameter $\lambda$ is a pure
consequence of the changing oscillator quantum number $Q$ of the
model.

A simple estimate of the strength of the subshell closure at $Z = 40$
can be obtained as follows. If there is a strong shell closure, a
discontinuity in $\lambda$ should also be observed when passing the
shell closure without changing the node number (as in the case around
$Z = 50$ discussed above). For this purpose the open symbols in
Fig.~\ref{fig:lamz} have been calculated at $Z = 38$ ($^{92}$Zr =
$^{88}$Sr $\otimes$ \al ) with $Q = 16$ (instead of $Q = 14$) and at
$Z = 40$ ($^{94}$Mo = $^{90}$Zr $\otimes$ \al ) with $Q = 14$ (instead
of $Q = 16$). From the comparison with neighboring nuclei it is
evident that there is no strong shell closure at $Z = 40$ because the
variation of $\lambda$ is very smooth. The physical properties of \al
-cluster states are thus not affected by the subshell
closure at $Z = 40$.

\begin{figure}[htb]
\includegraphics[ bb = 70 65 470 380, width = 80 mm, clip]{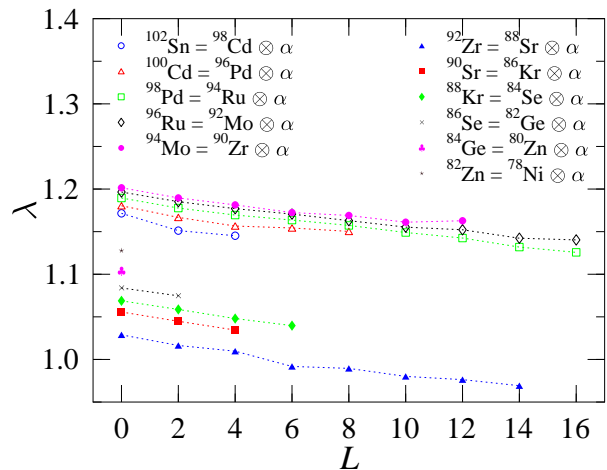}
\caption{
  \label{fig:laml} 
  Potential strength parameter $\lambda$ as a function of angular
  momentum $L$ for the ground state rotational bands in ($N$=50)
  $\otimes$ \al\ nuclei. For all nuclei $\lambda$ is slightly
  decreasing with increasing $L$. 
}
\end{figure}

For the example of $^{96}$Ru = $^{92}$Mo $\otimes$ \al\ it has already
been shown in Fig.~\ref{fig:exc} that the folding potential is able to
generate a rotational band. But the calculated excitation energies are
lower than the experimental energies. A small readjustment of the
potential strength parameter $\lambda$ for each state 
of the ground state rotational band is required to
obtain the correct excitation energies. This readjustment procedure
has been done for all nuclei under study. The result is shown in
Fig.~\ref{fig:laml}.

For all nuclei under study the same behavior is found. The potential
strength has to be reduced by less than 1\,\% for neighboring levels
of the rotational band (e.g., between $0^+$ and $2^+$ states or
between $6^+$ and $8^+$ states). The total change of the strength
parameter $\lambda$ within a rotational band, i.e.\ the change between
the $0^+$ ground state and the $14^+$ or $16^+$ state is about 5\,\%
in all nuclei under study.

\begin{figure}[htb]
\includegraphics[ bb = 55 65 470 380, width = 80 mm, clip]{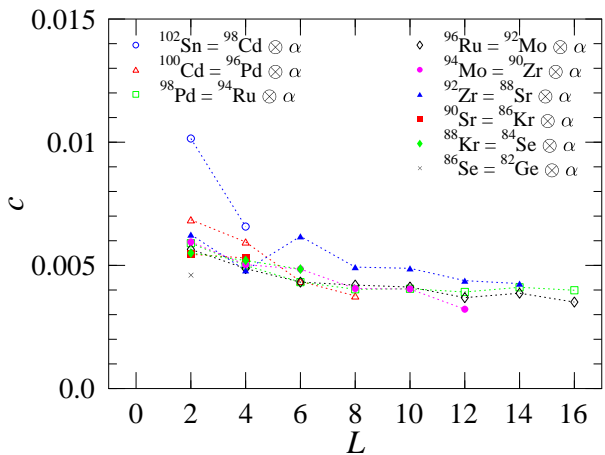}
\caption{
  \label{fig:lamc} 
  Variation of the potential strength parameter $\lambda$ according to
  Eq.~(\ref{eq:lampar}): the parameter $c$ is extracted from the
  previous Fig.~\ref{fig:laml}. By definition, $c = 0$ for $L = 0$
  (not shown).
}
\end{figure}

It has been suggested in \cite{Mic98,Ohk95} to parametrize the
variation in the potential strength parameter $\lambda$ by
\begin{equation}
\lambda(L) = \lambda(L=0) - c \times L
\label{eq:lampar}
\end{equation}
where the $\lambda(L=0)$ values are the required strength parameters
for the ground state binding energies (see Fig.~\ref{fig:lamz}). As
already pointed out above, 
the variations in $\lambda$ are very small. Thus the parameter $c$
is extremely small, and variations in $c$ are hardly visible in
Fig.~\ref{fig:laml}. Therefore in Fig.~\ref{fig:lamc} the
parameter $c$ is shown separately; it is extracted from all data in
Fig.~\ref{fig:laml}. The potential variation parameter $c$ is almost
constant for all nuclei and all angular momenta. There is a weak
tendency of smaller $c$ values for states with higher angular momenta
at higher excitation energies.

\begin{figure}[htb]
\includegraphics[ bb = 45 65 470 265, width = 80 mm, clip]{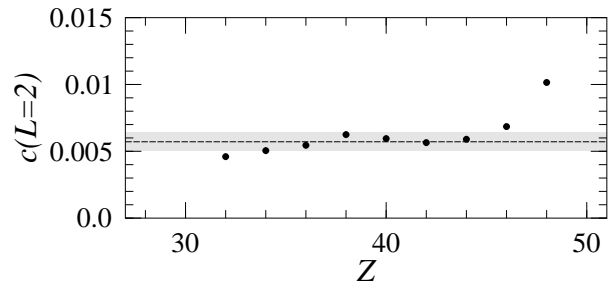}
\caption{
  \label{fig:cz} 
  The parameter $c(L=2)$ is shown as a function of $Z$. An average
  value of $c = 0.0057 \pm 0.0007$ (indicated by the shaded area) is
  found with the exception of $c = 0.0102$ for $Z = 48$ which results
  from the $Z_C = 50$ shell closure (see text).  
}
\end{figure}

For the first excited $2^+$ state the parameter $c(L=2)$ takes values
between about 0.005 and 0.007. This result is illustrated in
Fig.~\ref{fig:cz} where $c(L=2)$ is plotted against the proton number
$Z$. An average value of $c(L=2) = 0.0057 \pm 0.0007$ is found for the
nuclei between $Z = 32$ and $Z = 46$. An exceptionally large value of
$c = 0.0102$ is found for $Z = 48$, i.e., for the first $2^+$ state in
$^{102}$Sn = $^{98}$Cd $\otimes$ \al . Again, such an exceptional
behavior is a signature of a shell closure. The first excited state of
the semi-magic $Z_C = 50$ nucleus $^{102}$Sn is located at a relatively high
excitation energy of $E_x = 1472$\,keV \cite{ENSDF}. As a consequence,
a relatively small potential strength parameter $\lambda$ and a
relatively strong variation parameter $c(L=2)$ are required for the
correct description of the excitation energy of this $2^+$
state. Interestingly, the excitation energy of the $4^+$ state in
$^{102}$Sn can be calculated using a much smaller and almost regular
value for $c(L=4)= 0.0066$ (see Fig.~\ref{fig:lamc}). 

Experimental data for transition strengths are available for nuclei
close to stability whereas almost no transition strengths have been
measured for nuclei far from the valley of stability. In Table
\ref{tab:BE} the experimentally available transition strength data
\cite{ENSDF} are compared to the calculated values in the \al -cluster
model. Additionally, $B(E2)$ values for $2^+ \rightarrow 0^+$
transitions from the first excited $2^+$ state to the $0^+$ ground
state are listed for all nuclei under study.
The results in Table \ref{tab:BE} have been calculated without
effective charge. In general, the theoretical results deviate by less
than a factor of two from the experimental values. 
\begin{table}
  \caption{\label{tab:BE} Experimental \cite{ENSDF} and calculated
    transition strengths $B(E{\cal{L}})$ for \al -cluster states above
    $N=50$ nuclei. All transition strengths are given in Weisskopf
    units. The result for $^{104}$Te has been taken from \cite{Mohr07}.
}
\begin{center}
\begin{tabular}{c@{\,=\,}c@{\,$\otimes$ \al\ ~ }cc@{$\rightarrow$}ccc}
\multicolumn{3}{c}{nucleus}
& $L_i$
& $L_f$
& ~$B(E2)_{\rm{calc}}$~
& $B(E2)_{\rm{exp}}$~ \\
\hline
$^{82}$Zn & $^{78}$Ni & & $2^+$ & $0^+$ & 7.1 & $-$ \\
$^{84}$Ge & $^{80}$Zn & & $2^+$ & $0^+$ & 7.3 & $-$ \\
$^{86}$Se & $^{82}$Ge & & $2^+$ & $0^+$ & 7.4 & $-$ \\ 
$^{88}$Kr & $^{84}$Se & & $2^+$ & $0^+$ & 7.4 & $-$ \\ 
$^{90}$Sr & $^{86}$Kr & & $2^+$ & $0^+$ & 7.4 & 8.4(24) \\ 
\multicolumn{3}{c}{~} & $4^+$ & $2^+$ & 10.2 & 5.1(9) \\ 
$^{92}$Zr & $^{88}$Sr & & $2^+$ & $0^+$ & 7.8 & 6.4(6) \\ 
\multicolumn{3}{c}{~} & $4^+$ & $2^+$ & 10.6 & 4.04(12) \\ 
\multicolumn{3}{c}{~} & $8^+$ & $6^+$ & 9.1 & 3.59(22) \\ 
$^{94}$Mo & $^{90}$Zr & & $2^+$ & $0^+$ & 8.9 & 16.0(4) \\ 
\multicolumn{3}{c}{~} & $4^+$ & $2^+$ & 12.4 & 26(4) \\ 
$^{96}$Ru & $^{92}$Mo & & $2^+$ & $0^+$ & 8.7 & 18.0(6) \\ 
\multicolumn{3}{c}{~} & $4^+$ & $2^+$ & 12.1 & 21(3) \\ 
\multicolumn{3}{c}{~} & $6^+$ & $4^+$ & 12.1 & 12(8) \\ 
$^{98}$Pd & $^{94}$Ru & & $2^+$ & $0^+$ & 8.6 & $-$ \\ 
$^{100}$Cd & $^{96}$Pd & & $2^+$ & $0^+$ & 8.5 & $-$ \\ 
\multicolumn{3}{c}{~} & $8^+$ & $6^+$ & 10.5 & 0.0166(11) \\ 
$^{102}$Sn & $^{98}$Cd & & $2^+$ & $0^+$ & 8.5 & $-$ \\ 
$^{104}$Te & $^{100}$Sn & & $2^+$ & $0^+$ & 10.1 & $-$ \\ 
\end{tabular}
\end{center}
\end{table}

\section{\label{sec:disc}Discussion}
A very smooth variation of the potential strength parameter $\lambda$
and the variation parameter $c$ has been found for all nuclei under
study (see Figs.~\ref{fig:lamz} -- \ref{fig:cz}). This smooth behavior
can be used to predict up-to-now unknown properties. As an example, I
calculate the excitation energies of the first excited $2^+$ states of
$^{82}$Zn = $^{78}$Ni $\otimes$ \al\ and $^{84}$Ge = $^{80}$Zn
$\otimes$ \al . Properties of $^{104}$Te = $^{100}$Sn $\otimes$ \al\
have already been calculated in \cite{Mohr07}.

From the average value of $c = 0.0057$ for the $L = 2$ states one
finds $\lambda(L=2) = 1.0922$ for the first $2^+$ state in
$^{84}$Ge. The corresponding energy is $E_x = 877$\,keV. The
uncertainty of the potential variation parameter $c(L=2)$ translates
to an uncertainty of about 100\,keV for the excitation energy of the
$2^+$ state in $^{84}$Ge. A similar calculation for the first $2^+$
state in $^{82}$Zn leads to an excitation energy of $E_x = 880$\,keV,
again with an uncertainty of abot 100\,keV.

The excitation energy of the first excited $2^+$ state in $^{104}$Te
was estimated in \cite{Mohr07} as $E_x \approx 650$\,keV using values
of $c \approx 0.003 - 0.005$ from very few neighboring nuclei below
and above $^{104}$Te = $^{100}$Sn $\otimes$ \al . Using $c = 0.0057$
from this study of lighter $N=50$ nuclei, the revised excitation energy
is slightly higher: $E_x = 807$\,keV.

Besides the ground state band, odd-parity and higher-nodal bands have
been identified in lighter nuclei, e.g.\ $^{20}$Ne = $^{16}$O
$\otimes$ \al\ and $^{44}$Ti = $^{40}$Ca $\otimes$ \al\ (as reviewed
in \cite{Mic98}). The most successful experimental method has been \al
-transfer in the \rlid\ reaction \cite{Yam98}. Unfortunately, no
experimental \rlid\ data can be found for the nuclei under study in
\cite{ENSDF}. Such experiments are difficult because of the relatively
high Coulomb barrier for intermediate mass nuclei which leads to small
reaction cross sections and because of the increasing level density.
Furthermore, experimental studies using the
\rlid\ reaction are almost impossible for unstable $N = 50$
nuclei. Higher-nodal bands have also not been identified in an \al
-transfer experiment using the
$^{90}$Zr($^{16}$O,$^{12}$C$\gamma$)$^{94}$Mo reaction
\cite{Bohn72}.

As an example for higher-nodal bands, I calculate the excitation
energies of the band heads of the $Q = 17$ and $Q = 18$ bands in
$^{96}$Ru = $^{92}$Mo $\otimes$ \al\ and $^{94}$Mo = $^{90}$Zr
$\otimes$ \al . Using Eq.~(\ref{eq:lampar}), one finds $E_x \approx
7$\,MeV for the $1^-$, $Q = 17$ band heads and $E_x \approx 11$\,MeV
for the $0^+$, $Q = 18$ band heads in both nuclei. This finding is in
reasonable agreement with earlier estimates for $^{94}$Mo
\cite{Mic98,Mic00,Ohk95}.

In principle, the calculation of transition strengths from
Eq.~(\ref{eq:BE}) is straightforward. However, as can be seen from
Fig.~\ref{fig:BE}, the integrand in Eq.~(\ref{eq:BE}) consists of the
product of two oscillating wave functions. Thus, the integrand also
oscillates, and the integral depends sensitively on the zeroes of the
wave functions, i.e.\ the radial location of the nodes. This is
particularly the case for wave functions with few nodes where positive
and negative regions of the integrand in Eq.~(\ref{eq:BE}) may cancel
each other. For $E2$, $2^+ \rightarrow 0^+$ transitions the dominating
contribution comes from the nuclear surface (see Fig.~\ref{fig:BE}),
and the calculated $B(E2)$ value is not extremely sensitive to the
nodes of the wave functions and does not depend very sensitively on
the underlying potential.

In general, the calculated $B(E2)$ values do not deviate by more than
a factor of two from the experimental values (see Table
\ref{tab:BE}). This must be considered as a quite satisfactory result
because the nuclear structure of $A \approx 80-100$ nuclei is much
more complex than the simple \al -cluster description. On the other
hand, the reasonable agreement between calculated and experimental
$B(E2)$ values confirms that \al -clustering is still an important
feature for intermediate mass nuclei which has already been pointed
out earlier (e.g.\ \cite{Mic98,Ohk95,Buck95}).

There is one striking deviation between calculated and experimental
$B(E2)$ values in Table \ref{tab:BE}: the strength of the $E2$
transition from the $8^+$ isomer at $E_x = 2548$\,keV to the $6^+$
state at $E_x = 2095$\,keV in $^{100}$Cd is overestimated by a factor
of more than 500. It is a very special feature of the $^{100}$Cd
nucleus that two $4^+$ levels, two $6^+$ levels, and two $8^+$ levels
are located very close to each other: $E_x(4^+) = 1799$\,keV and
2046\,keV, $E_x(6^+) = 2095$\,keV and 2458\,keV, $E_x(8^+) =
2548$\,keV and 3200\,keV \cite{ENSDF}. The assignment of each lower
state to the ground state band in \cite{ENSDF} seems to be at least
questionable because of the extremely low transition strength from the
$8^+$ isomer to the $6^+$ state at $E_x = 2095$\,keV. A larger
strength of $B(E2) = 1.8(8)$\,W.u.\ has been found for the transition
from the $8^+$ isomer to the $6^+$ state at $E_x =
2458$\,keV. Unfortunately, no transition strength is known for the
decay of the second $8^+$ state at $E_x = 3200$\,keV which decays only
to the lower $6^+$ state at $E_x = 2095$\,keV. Summarizing the above,
a clear band assignment is not possible for $^{100}$Cd. Also mixing
may occur between states with the same quantum number $J^\pi$. Further
evidence for inconsistencies in the band assignment and/or mixing can
be read from Fig.~\ref{fig:lamc}. Here one finds relatively large $c$
values for $L=2$ and $L=4$; contrary, the $c$ values for $L=6$ and
$L=8$ are relatively small.

The present study is restricted to semi-magic $N=50$ even-even
nuclei. An extension to semi-magic $N=50$ even-odd nuclei is
complicated by the additional spin $I$ of the even-odd $N=50$ core which
couples to the angular momentum $L$ of the \al -particle and leads to
multiplets of \al -cluster states. A first attempt for $^{93}$Nb =
$^{89}$Y $\otimes$ \al\ has been made together with a study of the
$^{89}$Y\raa $^{89}$Y scattering cross section \cite{Kiss08}. Here I
briefly summarize the results of \cite{Kiss08}. $^{89}$Y
has $I^\pi = 1/2^-$ corresponding to neighboring $^{90}$Zr with a
proton hole in the $p_{1/2}$ subshell. \al -cluster states with $L =
0$, $J^\pi = 1/2^-$ and $L = 2$, $J^\pi = 3/2^-$, $5/2^-$ have been
clearly identified in $^{93}$Nb. The systematics of the potential
parameters strengthens a reassignment of $J^\pi$ of the state at $E_x
= 1500$\,keV in $^{93}$Nb. Whereas $J^\pi = 7/2$ is found in
\cite{ENSDF}, a recent experiment \cite{Orce07} has found strong
evidence for $J^\pi = 9/2^-$ and measured a transition strength of
$26.4^{+9.7}_{-6.2}$\,W.u.\ for the transition to the $L = 2$, $J^\pi
= 5/2^-$ \al -cluster state at $E_x = 810$\,keV. These experimental
data \cite{Orce07} confirm that the state at $E_x = 1500$\,keV is the
$L = 4$, $J^\pi = 9/2^-$ \al -cluster state in $^{93}$Nb = $^{89}$Y
$\otimes$ \al .

The folding potential has been calculated throughout this work using a
simplistic scaling of the radius parameter of the density of the core
nucleus with $r \sim A^{1/3}$. This global parametrization for the density
has been applied successfully in a broad mass range
\cite{Mohr06,Mohr07}. The obtained results do not depend sensitvely on
the chosen radius parameter. For a study of this sensitivity, I have
increased the radius parameter of the density of $^{92}$Mo strongly by
10\,\%. Because the other ingredients of the folding procedure, i.e.\
the \al -particle density and the interaction, remain unaffected, the
root-mean-square radius of the potential changes by only about
5\,\%. Although the absolute values of the potential strength change
(standard potential: $\lambda = 1.1965$ and $1.1852$ for the $0^+$ and
$2^+$ states in $^{96}$Ru; increased radius potential: $\lambda =
1.2296$ and $1.2162$), the variation $c$ of the potential strength in
Eq.~(\ref{eq:lampar}) remains very small and changes from 0.0057 to
0.0067 (see also Figs.~\ref{fig:lamc} and \ref{fig:cz}). The
calculated $B(E{\cal{L}})$ value for the $2^+ \rightarrow 0^+$
transition in $^{96}$Ru increases by about 20\,\% using the potential
with the larger radius. These relatively small changes of the results
indicate that the simplistic scaling of the radius parameter in the
density parametrization can be used for extrapolations to unstable
nuclei with reasonable accuracy.

\section{\label{sec:conc}Conclusions}
The \al -cluster model has been applied successfully to intermediate
mass nuclei above the $N=50$ shell closure between $^{82}$Zn =
$^{78}$Ni $\otimes$ \al\ and $^{104}$Te = $^{100}$Sn $\otimes$ \al .
The underlying double-folding potentials show a systematic and very
smooth variation which has been studied in detail. This behavior
allows extrapolations with relatively small uncertainties.

The present study is restricted to even-even nuclei. It may be
extended to even-odd nuclei although additional complications will
arise from the spin of the even-odd $N=50$ core which leads to
multiplets of \al -cluster states for each angular momentum $L > 0$.

Transition strengths for $E2$ transitions have been calculated and are
in rough agreement with experimental results. A significant deviation
for $^{100}$Cd may be the result of an inconsistent band assignment in
\cite{ENSDF} and/or mixing.

The results of the present study confirm that \al -clustering is an important
feature in intermediate mass nuclei. Experimental data from \al\
transfer reactions, in particular \rlid\ data, are urgently needed to
verify the theoretical predictions and to identify higher-nodal bands
in intermediate mass nuclei.

%
%

\end{document}